\documentclass[aps,prb,twocolumn,groupedaddress]{revtex4-1}

\usepackage{xcolor}
\usepackage{hyperref}
\usepackage{graphicx}
\usepackage{amsmath}
\usepackage{amsfonts}

\newcommand{\grating}{\texorpdfstring{Si\textsubscript{3}N\textsubscript{4}}{Si 3 N 4} }

\hypersetup{
    colorlinks = true,
    citecolor = blue,
    linkcolor = blue,
    urlcolor = blue
}

\begin{document}

%Title of paper
\title{A semi-analytical approach for the characterization of ordered 3D~nano structures using grazing-incidence X-ray fluorescence}

\author{K.V.~Nikolaev}
\email[]{k.nikolaev@protonmail.com}
\affiliation{MESA+ Institute for Nanotechnology, University of Twente, The Netherlands}

\author{V.~Soltwisch}
\affiliation{Physikalisch-Technische Bundesanstalt, Abbestr. 2-12, 10587 Berlin, Germany}

\author{P.~H\"onicke}
\affiliation{Physikalisch-Technische Bundesanstalt, Abbestr. 2-12, 10587 Berlin, Germany}

\author{F.~Scholze}
\affiliation{Physikalisch-Technische Bundesanstalt, Abbestr. 2-12, 10587 Berlin, Germany}

\author{J.~de~la~Rie}
\affiliation{MESA+ Institute for Nanotechnology, University of Twente, The Netherlands}

\author{S.N.~Yakunin}
\affiliation{NRC Kurchatov Institute, Moscow, 123182, Russia}

\author{I.A.~Makhotkin}
\affiliation{MESA+ Institute for Nanotechnology, University of Twente, The Netherlands}

\author{R.W.E.~van~de~Kruijs}
\affiliation{MESA+ Institute for Nanotechnology, University of Twente, The Netherlands}

\author{F.~Bijkerk}
\affiliation{MESA+ Institute for Nanotechnology, University of Twente, The Netherlands}

\begin{abstract}
    Following the recent demonstration of grazing-incidence X-ray fluorescence (GIXRF) based characterization of the 3D atomic distribution of different elements and dimensional parameters of periodic nanoscale structures,
    this work presents
    a new computational scheme for the simulation of the angular dependent fluorescence intensities from such periodic 2D and 3D nanoscale structures.
    The computational scheme is based on the dynamical diffraction theory in many-beam approximation, which allows to derive a semi-analytical solution to the Sherman equation in a linear-algebraic form.
    The computational scheme has been used to analyze recently published GIXRF data measured on 2D \grating lamellar gratings,
    as well as on periodically structured 3D Cr nano pillars.
    Both the dimensional and structural parameters of these nanostructures have been reconstructed by fitting numeric simulations to the experimental GIXRF data.
    Obtained results  show good agreement with nominal parameters used in the manufacturing of the structures, as well as with reconstructed parameters based on the previously published finite element method simulations, in case of the \grating grating.
\end{abstract}


\maketitle

\section{Introduction}

    Achievements in the field of science and technology related to the manufacturing of nanoscale devices are usually associated with the systematic decrease of the characteristic sizes of the structures within such devices.
    Such a decrease in characteristic sizes can lead to a strong performance dependency  to minor variations in the device structure, its geometry and the elemental composition of different elements inside the structure.
    Prominent examples of such nanoscale device structures can be found in the microelectronic industry~\cite{markov2014limits,buitrago2016high}.
    Understanding and improving the performance of such devices therefore requires the use of nanometrology techniques which, at best, are capable
    to reconstruct the geometry of the structure and the three dimensional atomic concentration distributions of different elements.
    Such element selective analysis can be performed using grazing-incidence X-ray fluorescence (GIXRF) \cite{soltwisch2018element, andrle2019grazing}.
    GIXRF is based on the X-ray standing wave (XSW) which is excited due to the interference between incident and reflected radiation. Its position and angle dependent amplitude can substantially modulate the GIXRF intensities of an element depending on its location within the nanostructure. By varying the angle of incidence and/or incident photon energy, the location of the XSW field nodes and anti-nodes can be varied inside the nanostructure. Consequently, the emission of fluorescence radiation depends on the incident angle and the incident photon energy, as well as on the spatial distribution of the fluorescent atoms.

    Measurement procedures and data analysis for one dimensional
    depth distributions of fluorescent atoms have been well developed~\cite{zegenhagen2013xsw}
    and implemented for the study of epitaxial layers~\cite{kroger2011normal},
    multilayers~\cite{yakunin2014model},
    Langmuir-Blodgett films~\cite{novikova2003total} and shallow ion implant profiles \cite{P.Hoenicke2009},
    among others.
    However, if nanoscale devices, e.g. light-trapping structures in solar cells~\cite{kroger2011normal},
    field emitter arrays~\cite{fletcher2013field} and
    nanorods~\cite{malerba2015vertical} are to be characterized, the calculation of the XSW is more complex.
    The depth atomic distribution profiles of such structures can still be analyzed in the framework of conventional 1D XSW method with use of the effective layer approximation~\cite{kennedy1999oxidation}.
    In this approximation, the atomic concentration distribution is averaged along the lateral directions.
    But this approach does not take into account the diffraction on the lateral structures of the sample and is therefore only applicable in case of randomly distributed objects.
    Inherently, the information about lateral distribution is lost within the effective layer approach, and the effective atomic concentration profile can never fully explain the properties of such 2D or 3D devices.

    In recent works~\cite{soltwisch2018element, dialameh2018development},
    the sensitivity of GIXRF to the lateral distribution of atomic concentration in 2D and 3D structures of periodically arranged gratings and nanocolumns has been experimentally demonstrated.
    To achieve such sensitivity, a new experimental scheme has been employed, where measurements are done under different grazing incidence and azimuthal orientation angles.
    The optical matrix method~\cite{gibaud00x} used for the analysis in conventional XSW~\cite{yakunin2014model}
    does not allow analysis of the lateral distribution of atomic concentration.

    This problem of GIXRF data analysis for well-ordered structures has been addressed in~\cite{soltwisch2018element}, where the 2D structure of a lamellar \grating grating has been analyzed.
    The experimentally measured GIXRF curves were analyzed by solving the Maxwell's equations by means of a finite-element method (FEM)~\cite{pomplun2007adaptive}.
    However applicability of FEM is limited due to its high demand in computational effort.
    It quickly increases with the increase of the incident photon energy, the size and the dimensionality of the structure.
    The FEM simulations for the experiments on the 3D Cr nanocolumns published in~\cite{dialameh2018development} for instance are practically irrealizable.

    Thus, in this study we provide an alternative approach for the calculation of the XSW field intensities within regular nanostructures by deriving semi-analytic equations based on the dynamical diffraction theory.
    We derive the solution of the Sherman equation~\cite{sherman1955theoretical, P.Hoenicke2009} for the GIXRF intensity induced by XSW in the 3D periodic structure in linear-algebraic form.
    In order to test the new computational scheme,
    we perform numerical simulations for the same 2D lamellar grating as published in~\cite{soltwisch2018element} and compare them with the results of FEM simulations and measurements.
    The semi-analytical nature of the derived equations allowed us to strongly reduce the computational effort,
    and to perform analysis of GIXRF also from a 3D nanostructured surface for the first time using the experimental data previously published in~\cite{dialameh2018development}.

\section{Theory \label{sec:theory}}

    In \hyperref[sec:theory]{Sections~\ref{sec:mbddt}} -- \hyperref[sec:boundary]{\ref{sec:boundary}} we consider the theoretical background of the dynamical diffraction theory in many beam approximation (MBDDT)~\cite{mikulik1999x}
    (in literature also refereed to as the rigorous coupled-wave analysis~\cite{chateau94diffraction}).
    In \autoref{sec:yield} we derive the solution  of the Shermann equation in linear-algebraic form, which will further allow us to calculate GIXRF intensities of 2D and 3D structures.

    \begin{figure*}
        \center\includegraphics[width = 1.5\columnwidth]{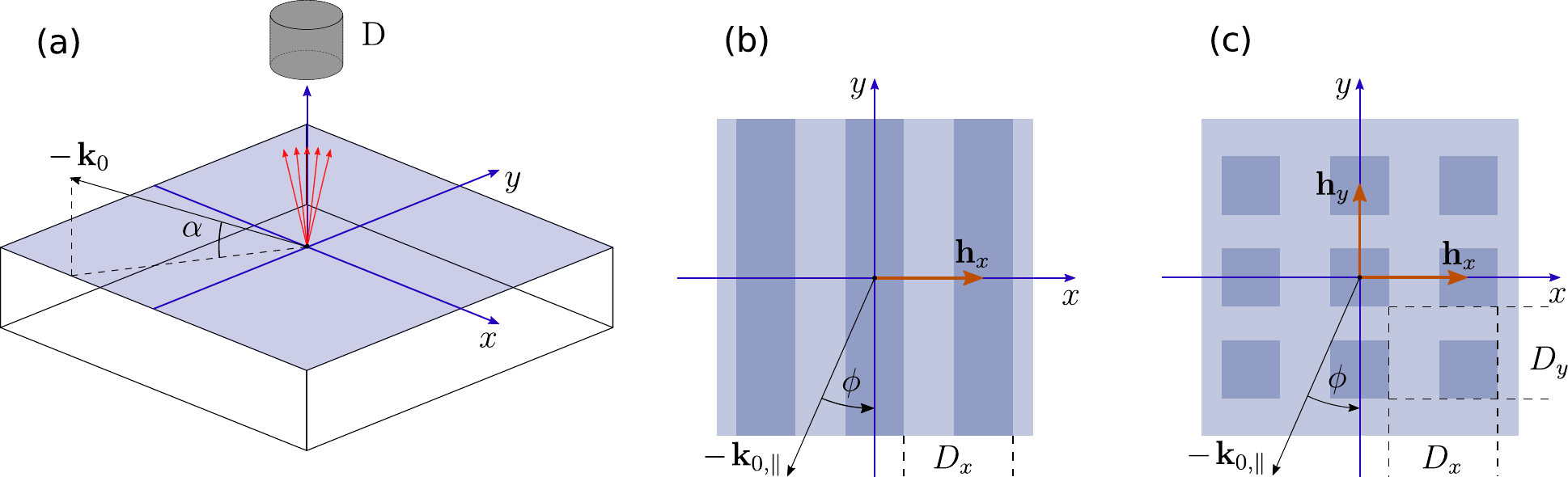}
        \caption{\label{fig:sketch}
                a) Sketch of the experimental geometry of GIXRF; $D$ -- energy-dispersive silicon drift detector,
                $\alpha$ -- angle of incidence,
                $\mathbf{k}_0$ -- wave vector of the incident beam.
                b) Sketch of a typical 2D periodic structure, with the azimuthal rotation angle $\phi$.
                c) Sketch of a typical 3D periodic structure.
                }
    \end{figure*}

\subsection{Many beam dynamical diffraction theory \label{sec:mbddt}}

    Experimental geometry used in~\cite{dialameh2018development,soltwisch2018element} for the GIXRF measurments is shown in \hyperref[fig:sketch]{Fig.~\ref{fig:sketch}a}.
    An X-ray beam impinges onto a sample surface under the grazing incidence angle $\alpha$ and azimuthal angle $\phi$.
    The excited fluorescence emission is measured using an energy-dispersive silicon drift detector $D$.
    To simulate the fluorescence intensity from the sample,
    the near field distribution within the nanostructure must be calculated.
    The problem of near field (NF) calculation is formulated by the Helmholtz equation:
    \begin{equation}
        (\Delta+k_0^2)E(\mathbf{r}) = -k_0^2\chi(\mathbf{r})E(\mathbf{r}).
        \label{eq:helmholtz}
    \end{equation}
    Here, for simplicity we consider the Helmholtz equation in a scalar approximation,
    as effect of polarization is negligible in grazing-incidence geometry in the X-ray spectral range;
    $E(\mathbf r)$ is the electric field,
    the sample structure is represented by the dielectric susceptibility function $\chi(\mathbf{r})$
    and $k_0 = 2\pi/\lambda$ is the wave number of the incident beam with the wavelength $\lambda$.
    The Helmholtz equation can be solved using the finite element method (FEM), kinematical diffraction theory or dynamical diffraction theory.
    With FEM being computationally challenging, and kinematical theory is not sufficiently precise under grazing incidence conditions~\cite{mikulik1999x},
    we further consider the dynamical diffraction theory.
    Furthermore, to take into account the lateral structure of the sample one needs to consider the dynamical diffraction theory in many beam approximation (MBDDT).

    In the dynamical diffraction theory, \autoref{eq:helmholtz} is solved assuming that NF is represented as a Bloch wave:
    \begin{equation}
        E(\mathbf r) = \sum_{\mathbf h} E_{\mathbf h}(z) \exp(i{\mathbf k}_{\mathbf{h}\parallel} \cdot {\mathbf r}),
        \label{eq:bloch}
    \end{equation}
    and the structure is represented as the Fourier series:
    \begin{equation}
        \chi({\mathbf r}) = \sum_{\mathbf h} \chi_\mathbf{h} \exp(i\mathbf{h} \cdot {\mathbf r}),
        \label{eq:chi}
    \end{equation}
    where $\chi_{\mathbf{h}}$ is the Fourier component:
    \begin{equation}
        \chi_{\mathbf{h}} = \frac{1}{\Omega}\iint\chi(x,y)e^{-i\mathbf{h}\cdot\mathbf{r}}dS.
    \end{equation}
    Here integration is taken over the unit cell area $\Omega$, for the corresponding reciprocal space vector:
    \begin{equation}
        \mathbf{h}_{x,y} =  \frac{ 2\pi n_{x,y} }{ D_{x,y} } \mathbf{e}_{x,y},
    \end{equation}
    with order of diffraction index $n_{x,y}$ and
    $D_{x,y}$ the periods along $x$ or $y$ directions respectively.
    The parallel component of the wave vector of the $h$-th diffraction order ${\mathbf k}_{\mathbf{h}\parallel} = \mathbf{k}_{0\parallel} + \mathbf{h}$
    is translationally invariant along the $z$ direction;
    i.e. ${\mathbf k}_{\mathbf{h}\parallel}$ is constant across all medias in a layered system for given $h$,
    while the vertical component is generally different in each medium and defined with the spherical dispersion equation:
    \begin{equation}
        q_{\mathbf{h}z}^2 = k_0^2 (1+\chi_0) -  k^2_{\mathbf{h}\parallel}.
        \label{eq:spherical}
    \end{equation}
    This equation is derived assuming that diffraction scattering is an elastic process: $k_{\mathbf{h}} = (1+\chi_0)k_0$
    and assuming translational invariance of ${\mathbf k}_{\mathbf{h}\parallel}$ mentioned above.
    Finally, substituting \autoref{eq:bloch}, \autoref{eq:chi} and \autoref{eq:spherical} in \autoref{eq:helmholtz},
    considering a property of the Fourier components: $\chi_{\mathbf{g}}\exp(i\mathbf{h}\cdot \mathbf{r}) = \chi_{\mathbf g - \mathbf h}$,
    result in a system of inhomogeneous linear ordinary differential equations (ODE) of second order:
    \begin{equation}
        q_{\mathbf{h}z}^2 E_{\mathbf h}(z) + {\frac  {{\mathrm  {d^2}}}{{\mathrm  {d}}z^2}} E_{\mathbf h}(z) + k_0^2\sum_{\mathbf g \neq \mathbf h} E_{\mathbf g}(z) \chi_{\mathbf g - \mathbf h} = 0.
        \label{eq:system}
    \end{equation}
    The general solution of such a system of ODE is a linear combination of particular solutions of corresponding homogeneous ODEs,
    where the $n$-th particular solution has the form of a standing wave with amplitudes $T_n$ and $R_n$.
    Thus, the $\mathbf{h}$-th solution of~\autoref{eq:system} has the form:
    \begin{equation}
        E_{\mathbf h}(z) = \sum_n \left[ T_n\exp(-ik_{nz}z) + R_n\exp(ik_{nz}z) \right] E_{hn},
        \label{eq:nfz}
    \end{equation}
    with linear combination coefficients $E_{hn}$.
    Therefore, the distribution of the NF is defined with \autoref{eq:bloch} and \autoref{eq:nfz}.
    Thus, the problem of NF calculation is reduced to finding $k_{n,z}$, $E_{hn}$, $T_n$ and $R_n$.

\subsection{Characteristic equation \label{sec:eigen}}

    In this section we discuss the calculation of $k_{n,z}$ and $E_{hn}$.
    Variable $k_{n,z}$ has a physical meaning as the vertical component of the wavevector (see \autoref{eq:nfz}).
    It defines the phase of the standing wave in the structured layer.
    One can assume that $k_{n,z}$ is defined with spherical dispersion $k_{z} = q_{z}$,
    however under that assumption \autoref{eq:system} has no solutions.
    Therefore, values of $k_{n,z}$ deviate from spherical dispersion.
    To calculate the precise value of $k_{n,z}$ in the structured layer
    one can substitute \autoref{eq:nfz} in \autoref{eq:system}.
    The result is represented as the eigenvalues-eigenvectors problem:
    \begin{equation}
		\left( \mathbf A - k_{zn}^2 \mathbf I \right)
        \mathbf{E}_n = \mathbf{0},
		\label{eq:disper}
	\end{equation}
    where $k_{zn}^2$ is an eigenvalue of matrix $\mathbf A$ and
    $\mathbf{E}_n$  is an eigenvector composed of the coefficients $E_{hn}$ from \autoref{eq:nfz}:
    $\mathbf{E}_n = (\dots E_{-1,n}, E_{0,n}, E_{1,n} \dots)^T$;
    $\mathbf A$ is of the form:
	\begin{equation}
		\mathbf A =
		k_0^2\mathbf C - \mathbf{X}.
	\end{equation}
	Matrix $\mathbf C$ is the Toeplitz circulant matrix:
	\begin{equation}
		\mathbf C =
		\left[
		\begin{array}{lllll}
			\ddots & & & &  \\
			 & \chi_0 & \chi_{-1} & \chi_{-2} &  \\
			 & \chi_1 & \chi_0 & \chi_{-1} &  \\
			 & \chi_{2} & \chi_{1} & \chi_{0} &  \\
			 & & & & \ddots \\
		\end{array}
		\right],
    \label{eq:Toeplitz}
	\end{equation}
	and $\mathbf X$ is the diagonal matrix with diagonal $( \dots -k^2_{-1,\parallel},-k^2_{0,\parallel},-k^2_{1,\parallel} \dots )$.
    Circulant matrices have a remarkable property:
    with increasing circulant matrix size,
    its eigenvalues asymptotically approach the exact values for an infinite matrix~\cite{gray2006toeplitz}.
    Therefore, one can use a finite amount of Fourier components in \autoref{eq:chi} to approximate the exact solution of \autoref{eq:system}.
    Consider a set of $2N+1$ Fourier components $\{\chi_{-N}, \chi_{-N+1}, \dots ,\chi_0, \dots \chi_{N-1}, \chi_{N} \}$.
    These Fourier components constitutes a circulant matrix $\mathbf{C}$ of a size $\mathbf{C} \in \mathbb{C}^{M \times M}$, where $M = N+1$.
    Solving the characteristic \autoref{eq:disper} will give $M$ eigenvalue-eigenvector pairs.

\subsection{Boundary conditions \label{sec:boundary}}

    In this section we calculate the transmission $T_n$ and reflection $R_n$ amplitudes.
    Consider a sample as a stratified medium, consisting of layers.
    $T_n$ and $R_n$ are calculated in each layer using continuity conditions of the electric field and its first derivative.
    The continuity conditions~\cite{born2013principles} for the $j$-th and $(j+1)$-th pair of layers can be written in a matrix form:
    \begin{equation}
        \mathbf{P}^{(j)}
        \left[
        \begin{array}{c}
            \mathbf{T}^{(j)} \\ \mathbf{R}^{(j)}
        \end{array}
        \right]
        =
        \mathbf{P}^{(j+1)}\mathbf{Q}^{(j+1)}
        \left[
        \begin{array}{c}
            \mathbf{T}^{(j+1)} \\ \mathbf{R}^{(j+1)}
        \end{array}
        \right].
        \label{eq:continuity}
    \end{equation}
    Here $\mathbf{T}$ and $\mathbf{R}$ are vectors composed of amplitudes $T_n$ and $R_n$:
    \begin{equation}
        T = (T_{-N/2}, T_{-N/2+1}, \dots ,T_0, \dots T_{N/2-1}, T_{N/2} )^T.
    \end{equation}
    \autoref{eq:continuity} links amplitudes $\mathbf{T}^{(j)}, \mathbf{R}^{(j)}$ at the interface between $(j-1)$-th and $j$-th layer,
    and amplitudes $\mathbf{T}^{(j+1)}$, $\mathbf{R}^{(j+1)}$ at the interface between $j$-th and $(j+1)$-th layer,
    Matrix $\mathbf P$ is the refraction matrix.
    For a structured layer it has a form:
    \begin{equation}
        \mathbf P =
        \left[
        \begin{array}{cc}
            \mathbf E & \mathbf E \\
            -\mathbf E  \mathbf k_z & \mathbf E  \mathbf k_z \\
        \end{array}
        \right],
    \end{equation}
    and for a homogeneous layer:
    \begin{equation}
        \mathbf P =
        \left[
        \begin{array}{cc}
            \mathbf I & \mathbf I \\
            -\mathbf k_z & \mathbf k_z \\
        \end{array}
        \right].
    \end{equation}
    Here, the matrix $\mathbf E$ is composed of columns of eigenvectors and matrix $\mathbf{k}_z$ is a diagonal matrix filled with $k_{z,n}$.
    Refraction matrix $\mathbf P$ is a $2\times 2$ block matrix, thus $\mathbf {P} \in \mathbb{C}^{2M\times 2M} $.
    Finally $\mathbf Q$ is the propagation matrix:
    \begin{equation}
        \mathbf Q =
        \left[
        \begin{array}{cc}
            \mathbf{Q}^+ & \mathbf 0 \\
            \mathbf 0 & \mathbf{Q}^- \\
        \end{array}
        \right],
    \end{equation}
    where $\mathbf{Q}^\pm$ are diagonal matrices with corresponding diagonals:
    \begin{equation}
        (\dots e^{\mp ik_{-h,z}d_j},e^{\mp ik_{0,z}d_j},e^{\mp ik_{h,z}d_j} \dots),
    \end{equation}
    where $d_j$ is the thickness of $j$-th layer.
    Although these equations can be used to calculate $\mathbf{T}_j$ and $\mathbf{R}_i$,
    solving \autoref{eq:continuity} might be problematic due to the poorly conditioned transmission matrix in case of sufficiently large thickness of the sample
    and/or in case of sufficiently high number of Fourier components used in the calculation.

\subsection{Numerical stability \label{sec:stability}}

\begin{figure}[h!]
        \center\includegraphics[width = 0.5\linewidth]{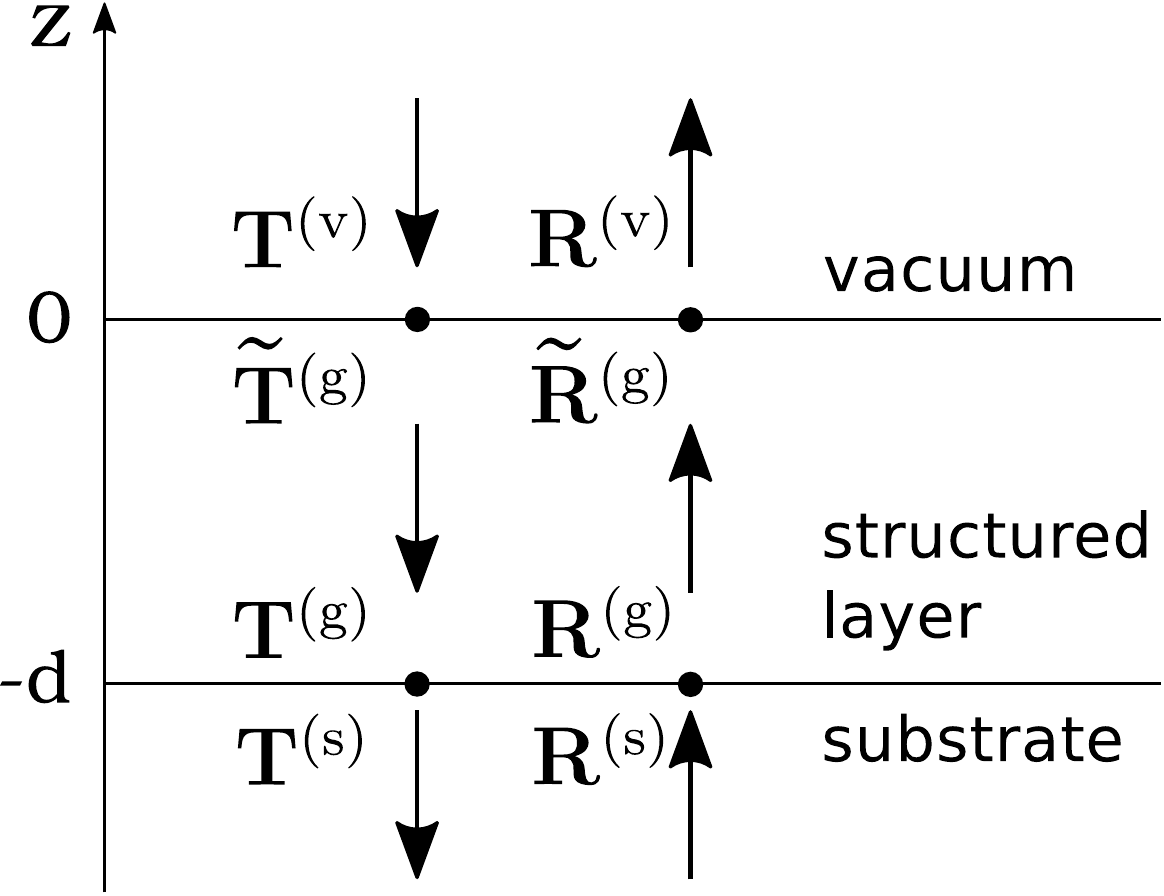}
        \caption{   Sketch of the three layer model.
                    Arrows schematically depicts the direction of propagation of the plane waves.
                    Amplitudes of the plane waves are assembled into $\mathbf{T}$ and $\mathbf{R}$ vectors.
                    $\widetilde{\mathbf{T}}$ and $\widetilde{\mathbf{R}}$ are amplitudes defined at the upper interface of the layer.
                    $\mathbf{T}$ and $\mathbf{R}$ are defined at the bottom interface of the layer.}
        \label{fig:field_sketch}
    \end{figure}

    The problem of numerical stability in the matrix formalism of dynamical diffraction theory was considered in~\cite{stepanov1998dynamical}.
    There the problem of numerical stability has been solved for the dynamical diffraction theory in two-beam approximation
    (only $\chi_{-1}$, $\chi_0$ and $\chi_1$ has been taken into account).
    It has been solved by dividing matrices in $2 \times 2$ block matrices and solving \autoref{eq:continuity} separately for each block matrix by using recurrent formula.
    Although the recurrent matrix equations in~\cite{stepanov1998dynamical} were derived for the two-beam case, they are generally applicable to the many-beam case.
    For brevity, we present these equations explicitly written for a three-layer system (see \autoref{fig:field_sketch}), which is relevant to the experimental data we will considered further.
    The continuity conditions for such a three-layer structure (vacuum -- structured layer -- substrate) are represented by the system of linear equations:
    \begin{gather}
        \left[
        \begin{array}{c}
            \mathbf{T}^{({\rm v})} \\ \mathbf{R}^{({\rm v})}
        \end{array}
        \right]
        =
        \left(\mathbf{P}^{({\rm v})}\right)^{-1} \mathbf{P}^{({\rm g})}\mathbf{Q}^{({\rm g})}
        \left[
        \begin{array}{c}
            \mathbf{T}^{({\rm g})} \\ \mathbf{R}^{({\rm g})}
        \end{array}
        \right];
        \nonumber\\
        \left[
        \begin{array}{c}
            \mathbf{T}^{({\rm g})} \\ \mathbf{R}^{({\rm g})}
        \end{array}
        \right]
        =
        \left(\mathbf{P}^{({\rm g})}\right)^{-1} \mathbf{P}^{({\rm s})}
        \left[
        \begin{array}{c}
            \mathbf{T}^{({\rm s})} \\ \mathbf{R}^{({\rm s})}
        \end{array}
        \right].
    \end{gather}
    One can rewrite that system as follows:
    \begin{gather}
        \left[
        \begin{array}{c}
            \mathbf{T}^{({\rm g})} \\ \mathbf{R}^{({\rm v})}
        \end{array}
        \right]
        = \mathbf{M}^{\rm vg}
        \left[
        \begin{array}{c}
            \mathbf{T}^{({\rm v})} \\ \mathbf{R}^{({\rm g})}
        \end{array}
        \right];
        \nonumber\\
        \left[
        \begin{array}{c}
            \mathbf{T}^{({\rm s})} \\ \mathbf{R}^{({\rm g})}
        \end{array}
        \right]
        = \mathbf{M}^{\rm gs}
        \left[
        \begin{array}{c}
            \mathbf{T}^{({\rm g})} \\ \mathbf{R}^{({\rm s})}
        \end{array}
        \right].
    \end{gather}
    Here matrix $\mathbf M$ has the form of a block matrix:
    \begin{equation}
        \mathbf{M} =
        \left[
        \begin{array}{cc}
            \mathbf{Q}^{-} \mathbf{V}_{11}^{-1} & -\mathbf{Q}^{-} \mathbf{V}_{11}^{-1} \mathbf{V}_{12}\mathbf{Q}^- \\
            \mathbf{V}_{21} \mathbf{V}_{11}^{-1} & \mathbf{V}_{22}\mathbf{Q}^{-} - \mathbf{V}_{21} \mathbf{V}_{11}^{-1} \mathbf{V}_{12}\mathbf{Q}^-
        \end{array}
        \right],
    \end{equation}
    where $\mathbf{V}_{ij}$ is a matrix element of $2 \times 2$ block matrix $\mathbf{V}^{\rm (vg,gs)} = \left(\mathbf{P}^{({\rm v,g})}\right)^{-1} \mathbf{P}^{({\rm g,s})}$
    Note that this equation does not include $\mathbf {Q}^+$ which elements are growing exponentially with respect to the thickness of the structured layer.
    Hence this matrix is numerically stable.
    Amplitudes $\mathbf T^{({\rm v})}$ represent the incident beam, therefore
    \begin{equation}
        \mathbf T^{({\rm v})} = (\dots 0,1,0 \dots)^T.
    \end{equation}
    Additionally, for a sufficiently thick substrate we can assume
    \begin{equation}
        \mathbf R^{({\rm s})} = (\dots 0,0,0 \dots)^T.
    \end{equation}
    Taking into account these considerations, we derive equations for amplitudes in the structured layer:
    \begin{equation}
        \mathbf R^{({\rm g})} = \left( \mathbf{I} - \mathbf{M}_{21}^{({\rm gs})}\mathbf{M}_{12}^{({\rm vg})} \right)^{-1} \mathbf{M}_{21}^{({\rm gs})} \mathbf{M}_{11}^{({\rm vg})}  \mathbf T^{({\rm v})},
        \label{eq:amp_rg}
    \end{equation}
    and:
    \begin{equation}
        \mathbf T^{({\rm g})} = \left( \mathbf{I} - \mathbf{M}_{12}^{({\rm vg})}\mathbf{M}_{21}^{({\rm gs})} \right)^{-1} \mathbf{M}_{11}^{({\rm vg})}  \mathbf T^{({\rm v})}.
    \end{equation}
    These amplitudes are calculated at the interface between the structured layer and the substrate (see \autoref{fig:field_sketch}).
    One can calculate amplitudes at the vacuum-structured layer interface using:
    \begin{equation}
        \left[
        \begin{array}{c}
            \widetilde{\mathbf{T}}^{({\rm g})} \\ \widetilde{\mathbf{R}}^{({\rm g})}
        \end{array}
        \right]
        = \left(\mathbf{P}^{({\rm g})}\right)^{-1} \mathbf{P}^{({\rm v})}
        \left[
        \begin{array}{c}
            \mathbf{T}^{({\rm v})} \\ \mathbf{R}^{({\rm v})}
        \end{array}
        \right].
        \label{eq:amp_tg}
    \end{equation}
    Finally, we need to rewrite \autoref{eq:nfz}:
    \begin{equation}
        E_{\mathbf h}(z) = \sum_n \left[ \widetilde T_n\exp(-ik_{nz}z) + R_n\exp(ik_{nz}[z+d]) \right] E_{hn}.
    \end{equation}
    Here, both exponents decrease with respect to the depth,
    providing numerical stability.

\subsection{X-ray fluorescence intensity \label{sec:yield}}

    The fluorescence intensity $Y$ can be calculated using the Sherman equation~\cite{sherman1955theoretical},
    adapted for GIXRF~\cite{P.Hoenicke2009}:
    \begin{equation}
        Y \propto G(\alpha) \iiint |E(\mathbf r)|^2 p(\mathbf r)  \exp(-\mu \rho z) d{\mathbf r},
        \label{eq:fluor_base}
    \end{equation}
    where $p(\mathbf r)$ describes the density distribution of fluorescent atoms in the structure,
    and $G(\alpha)$ is the geometrical factor~\cite{beckhoff2008reference, li2012geometrical, lubeck2013novel}.
    The integral is taken over the area of the elementary cell.
    The exponential term $\exp(-\mu \rho z)$ in \autoref{eq:fluor_base} takes into account the self absorption of emitted fluorescent photons.
    Here, $\mu$ is the absorption coefficient and
    $\rho$ is the effective density of the absorbing media.
    The integral in \autoref{eq:fluor_base} can be separated as follows
    (further, for brevity we do not explicitly write the multiplicative term $G(\alpha)$):
    \begin{multline}
        Y \propto \sum_{\mathbf{g},\mathbf{h}}
        \iint p(\mathbf{r}_\parallel) \exp(i[{\mathbf k}_{\mathbf{g}\parallel} - {\mathbf k}_{\mathbf{h}\parallel}^*] \cdot \mathbf{r}) dx dy
        \times
        \\
        \times
        \sum_{m,n} \int
        \left[ T_m\exp(-ik_{mz}z) + R_m\exp(ik_{mz}z]) \right]
        \times
        \\
        \times
        \left[ T_n^*\exp(ik^*_{nz}z) + R_n^*\exp(-ik^*_{nz}d) \right]
        \times
       \\
       \times
        E_{gm}E_{hn}^* \exp(-\mu \rho z) dz.
       \label{eq:big_integral}
    \end{multline}
    Such integral separation imposes a restriction on numeric density function:
    it must not be dependent on $z$ coordinate $p(\mathbf{r}) \equiv p(\mathbf{r}_\parallel)$;
    i.e. \autoref{eq:big_integral} can only be used in cases when fluorescent atoms are distributed homogeneously along the $z$ direction.
    Distribution in the $xy$~-plane can be arbitrary.
    In case of an inhomogeneous vertical distribution,
    one can discretize the structure along the $z$ direction as a stack of sublayers and calculate \autoref{eq:big_integral} for each sublayer.

    \autoref{eq:fluor_base} was rewritten in the form of \autoref{eq:big_integral},
    so it can be conveniently represented in a linear algebraic language.
    The fluorescence intensity can be expressed as the sum of matrix elements $ Y \propto \sum_{g,h}F_{gh}$ of the matrix:
    \begin{equation}
        \mathbf F = \mathbf \Phi \circ
        \left(
        \mathbf E  \mathbf \Psi  \mathbf E^*
        \right).
        \label{eq:la_gixrf}
    \end{equation}
    Here $\circ$ represent element-wise (Hadamard) multiplication.
    Elements of matrix $\mathbf \Phi$ have the form:
    \begin{equation}
        \Phi_{hg} \equiv \int\displaylimits_{-{D_x}/{2}}^{{D_x}/{2}} dx \int\displaylimits_{-{D_y}/{2}}^{{D_y}/{2}}  dy \; p(\mathbf{r}_\parallel) \exp(i[{\mathbf k}_{\mathbf{g}\parallel} - {\mathbf k}_{\mathbf{h}\parallel}^*] \cdot \mathbf{r}),
        \label{eq:Phi_matrix}
    \end{equation}
    and elements of matrix $\mathbf \Psi$ have the form:
    \begin{multline}
        \Psi_{mn} \equiv
        T_mT_n^*  U(-k_{zm}+k_{zn}^*)
        +T_mR_n^* U(-k_{zm}-k_{zn}^*)
        +
        \\
        +R_mT_n^* U(k_{zm}+k_{zn}^*)
        +R_mR_n^* U(k_{zm}-k_{zn}^*),
    \end{multline}
    where
    \begin{equation}
        U(q) \equiv \int\displaylimits_{-d}^0 \exp(iqz) \exp(-\mu \rho z) dz.
        \label{eq:psi_int}
    \end{equation}
    The $\mathbf \Phi$ matrix takes into account the distribution of fluorescent atoms and the electric field distribution in lateral direction and
    the $\mathbf \Psi$ matrix takes into account photon absorption and the electric field distribution in vertical direction.
    \autoref{eq:la_gixrf} allows to calculate the integral in \autoref{eq:fluor_base} analytically,
    which is much more computationally efficient compared to the numerical integration.

\section{Numerical simulations \label{sec:gixrf_results}}
\subsection{2D structure: \grating lamellar grating \label{sec:2Dgixrf}}

    \begin{figure}[t!]
        \center\includegraphics[width = \columnwidth]{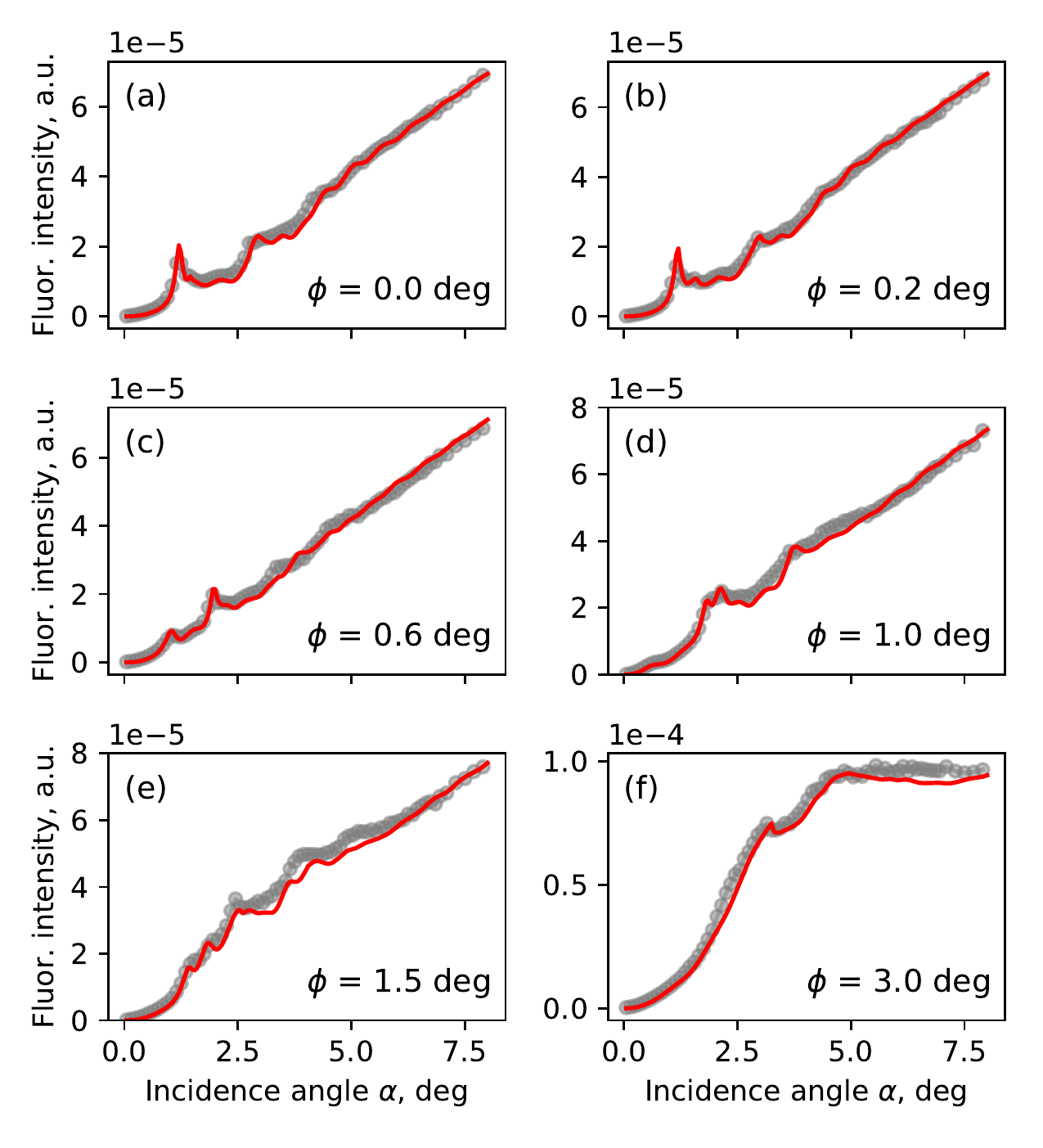}
        \caption{ N-K$\alpha$ GIXRF intensity, measured for various azimuthal orientation angles
                 (a) $\phi = 0^\circ$ - conical,
                 (b) $\phi = 0.2^\circ$,
                 (c) $\phi = 1^\circ$
                 and
                 (d) $\phi = 3^\circ$.
                  Red lines -- numerical simulation, gray markers -- experimental values.
                }
        \label{fig:2D_gixrf_curves}
    \end{figure}
    \begin{figure*}
        \center\includegraphics[width = 2\columnwidth]{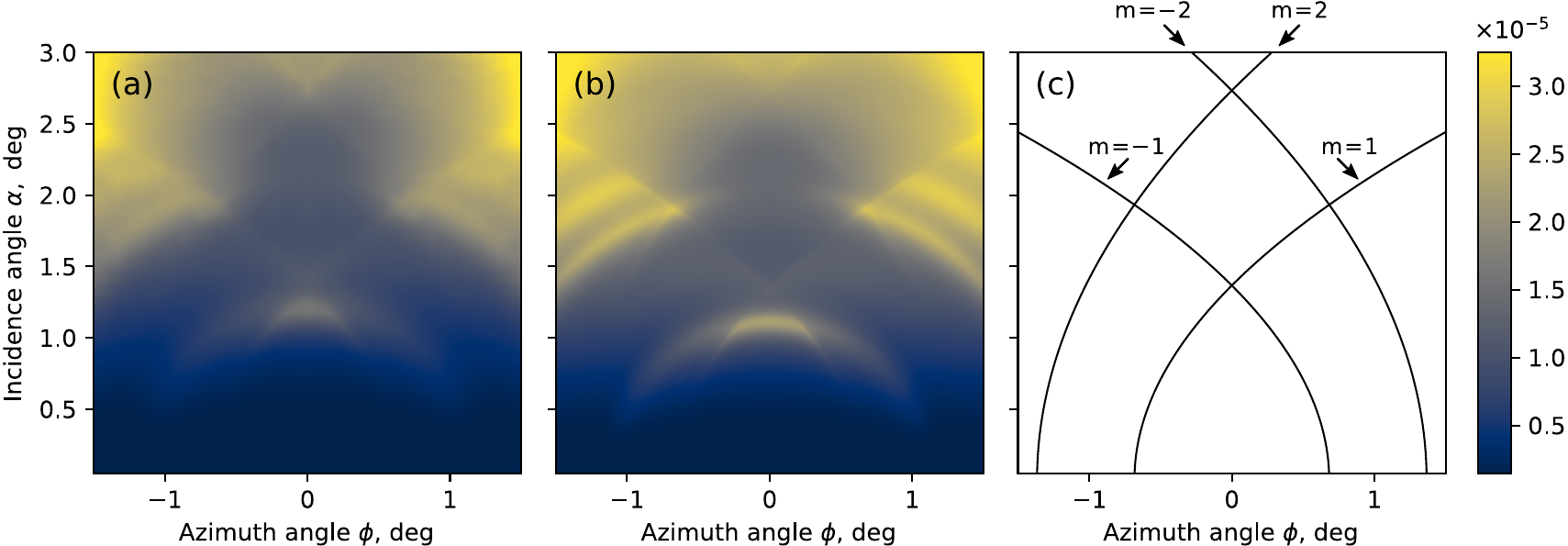}
        \caption{
                Comparison between the experimental GIXRF N-K$\alpha$ map
                (a) of the \grating lamelar grating measured with the incidence photon energy $E = 520~{\rm Ev}$
                and the simulated GIXRF map (b) based on a best fit model.
                (c) Resonant lines in GIXRF map for \grating grating structure,
                    caused by interference between reflected beam and m-th order of diffraction.
                }
        \label{fig:2Dmap}
    \end{figure*}

    Here, we consider a 2D lamellar \grating grating prepared using electron beam lithography.
    The original study with experimental data and numerical simulation of GIXRF intensity by means of FEM has been published by Soltwisch et.~al.~\cite{soltwisch2018element}
    The grating has a nominal period of $D_x = 100~{\rm nm}$,
    the thickness of the structured layer is $d = 90~{\rm nm}$
    and the line width is $40$~nm.

    The GIXRF measurements were carried out at the plane-grating monochromator (PGM) beamline \cite{senf1998plane}
    for undulator radiation at the PTB laboratory \cite{beckhoff2009quarter} of the BESSY II electron storage ring.
    % using a photon energy of $520~{\rm eV}$.
    A monochromatic excitation with a photon energy of $520~{\rm eV}$ was used.
    The GIXRF intensities were obtained for the N-K$\alpha$ fluorescence emission under various incidence angles $\alpha$
    and azimuthal sample orientation angles $\phi$ (see \hyperref[fig:sketch]{Fig.~\ref{fig:sketch}b});
    $\phi = 0^\circ$ corresponds to the conical orientation~\cite{goray2018rigorous} of the sample grating.
    The recorded spectra from the silicon drift detector were deconvoluted using detector response functions in order to isolate the fluorescence signal from N-K$\alpha$ from other spectral contributions. Further corrections, to take into account the detection efficiency and the geometrical factor (effective solid angle) were applied (see ref.~\cite{soltwisch2018element} for further details).

    Best-fit simulations obtained by sequential least squares optimization algorithm
    and experimental GIXRF data are shown in \autoref{fig:2D_gixrf_curves} for various azimuthal orientation angles $\phi$.
    For the simulation we use a simple box model,
    in which the grating lines are treated as an array of boxes on top of the substrate
    (see \hyperref[fig:sketch]{Fig.~\ref{fig:sketch}b}).
    Thus, the medium is divided into three areas: the vacuum, the structured layer in which the boxes are located, and the substrate.
    Within the box model,
    the sidewalls of the grating lines are considered to be parallel while
    the actual grating has a sidewall tilt angle.
    Based on the reconstruction in~\cite{soltwisch2018element},
    this angle is not greater than $\beta = 4^\circ$.
    In terms of the model it means that the Fourier transform in \autoref{eq:chi} is changing along $z$ axis.
    To compensate for that in the simulations,
    within one layer model
    averaged Fourier components have been used, i.e. $\langle \chi_h \rangle = \chi_h\exp(-h^2\sigma^2/2)$,
    with $\sigma$ defined as half the projection of the sidewall on the $x$ axis: $\sigma \equiv d\arctan{(\beta)}/2$.
    Best-fit line width (defined as the half-height width) is $D_l = 39~{\rm nm}$ and the best-fit sidewall tilt angle $\beta = 5^\circ$.

    Another feature of the actual sample that must be considered in the simulations is the effect of oxidation of surface and line edges.
    It affects the actual structure such that the concentration of fluorescent N atoms at the top part and at the line edges is strongly reduced.
    In the one layer model, oxidation of the surface can be effectively incorporated by changing the integration limits in \autoref{eq:psi_int},
    such that the integration in \autoref{eq:psi_int}
    is taken only over a range where fluorescent atoms are present.

    The best agreement with the experimental data was obtained with an effective surface layer thickness of $d_t = 3.3~{\rm nm}$ at the top of the lines.
    The best fit suggests that the N is not diluted at the line edges, since the reconstructed parameter of the effective edge thickness of the edges is $d_s = 0~{\rm nm} $.
    We note that this value $d_s$ is correlated with $\sigma$ used in averaging of the Fourier components, thus may be not representative.
    Also, note that these values only describes surface effects in terms of absence of fluorescent N atoms,
    ignoring the gradual change in stoichiometry throughout the surface and the edges.
    It also neglects the change of optical properties of the structure due to oxidation.
    Best-fit parameter of the grating height, excluding effective surface layer, is $d = 88.7$~nm.
    Average density of the line is $\rho_{{\rm Si}_3{\rm N}_4} = 2.8~{\rm g}/{\rm cm}^3$ and the density of the substrate is $\rho_{\rm Si} = 2.22~{\rm g}/{\rm cm}^3$.

    Best fit model and experimental data are qualitatively in good agreement.
    Qualitative agreement is also apparent on GIXRF intensity $(\alpha,\phi)$-maps shown in \autoref{fig:2Dmap}.
    A full set of 48 experimental GIXRF curves taken along different azimuthal angles $\phi$ (from $0^\circ$ untill $2^\circ$)
    was interpolated on a $(\alpha,\phi)$ grid (see  \hyperref[fig:2Dmap]{Fig.~\ref{fig:2Dmap}a}).
    The theoretical GIXRF map was calculated on the same $(\alpha,\phi)$ grid
    using best fit parameters from the data presented in \autoref{fig:2Dmap}.

    One can note a distinctive feature on the GIXRF map~-- resonant lines, which are visible both on the experimental data in \hyperref[fig:2Dmap]{Fig.~\ref{fig:2Dmap}a}
    and in the numerical simulations (see \hyperref[fig:2Dmap]{Fig.~\ref{fig:2Dmap}b}).
    As a visual aid to notice these lines one can refer to the sketch in \hyperref[fig:2Dmap]{Fig.~\ref{fig:2Dmap}c}.
    In \hyperref[fig:2Dmap]{Fig.~\ref{fig:2Dmap}c} the position of the resonant lines is marked with black contour lines.

    We assume that these lines are due to the interference between the reflected beam ($0$-th order of diffraction) and a diffracted beam
    ($m$-th order of diffraction).
    Therefore, the resonant lines must satisfy the Laue condition,
    which for this geometry can be formulated as $k_x^2+k_z^2 = (k_x+h)^2$.
    This formula geometrically corresponds to the Ewald sphere.
    For convenience we rewrite this equation in terms of the incidence and azimuthal angle:
    \begin{equation}
        \sin\phi = \dfrac{\sin^2\alpha-\gamma^2}{2\gamma\cos\alpha},
        \label{eq:ewald}
    \end{equation}
    where $\gamma = \lambda m/D_x$.
    The contour lines in \hyperref[fig:2Dmap]{Fig~\ref{fig:2Dmap}c} were calculated using this equation.
    Note that the resonant lines depend only on the lateral period of the structure $D_x$ and the wavelength $\lambda$ (see \autoref{eq:ewald},
    no other geometrical parameters are involved.
    Due to their explicit dependence on only the period of the structure,
    such lines might be used in the analysis of experimental data as a reference,
    to determine the lateral period of the structure,
    without needing a full structure reconstruction through model simulations.

\subsection{3D structure: Cr nanocolumns \label{sec:nanocolumns}}

    \begin{figure}[b!]
        \center\includegraphics[width = \columnwidth]{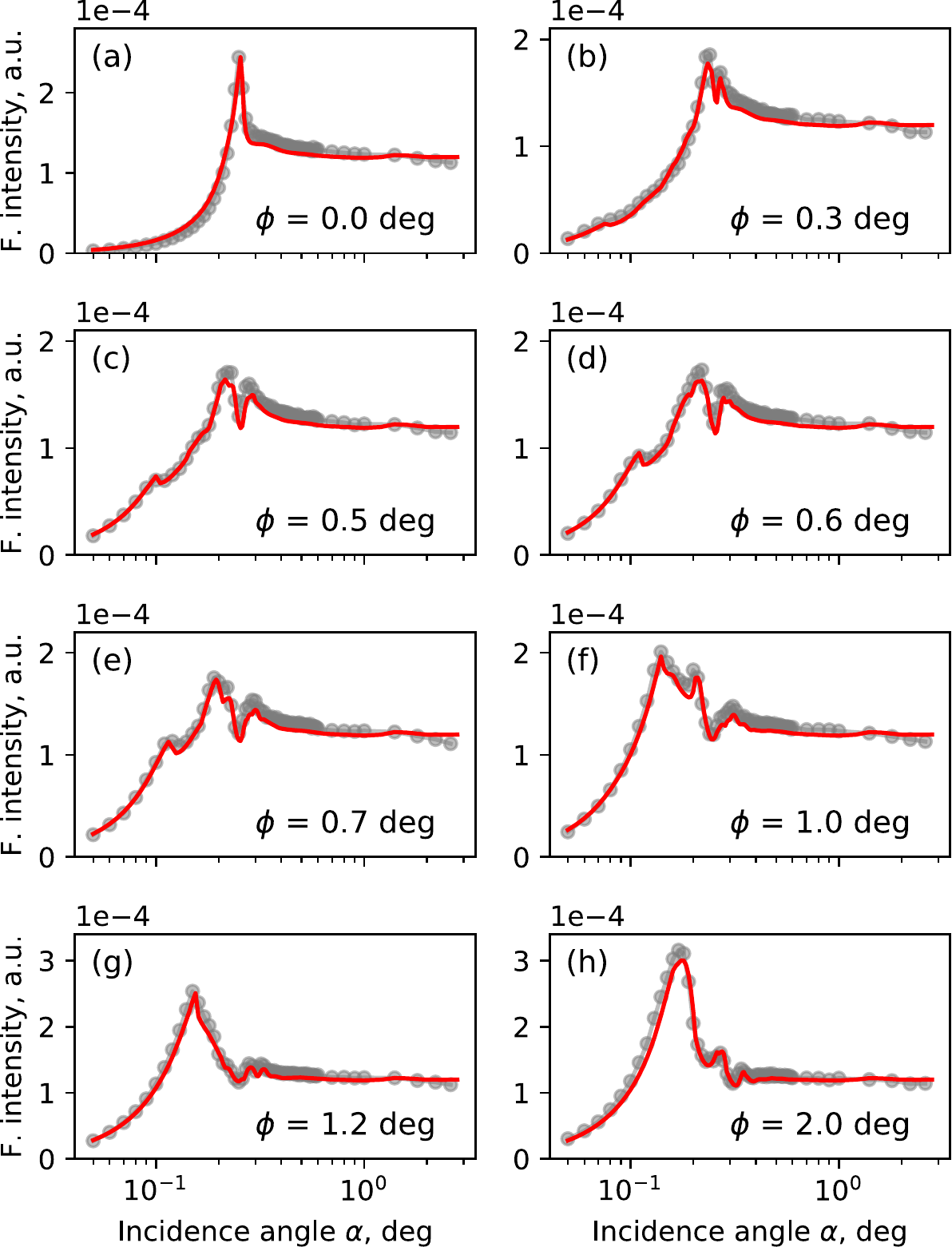}
        \caption{
                    Cr-K$\alpha$ GIXRF intensity curves, measured for various azimuthal orientation angles
                    (a) $\phi = 0^\circ$ - conical
                    (b) $\phi = 0.3^\circ$,
                    (c) $\phi = 0.5^\circ$,
                    (d) $\phi = 0.6^\circ$,
                    (e) $\phi = 0.7^\circ$,
                    (f) $\phi = 1.0^\circ$,
                    (g) $\phi = 1.2^\circ$,
                    and
                    (h) $\phi = 2.0^\circ$.
                    Red lines -- numerical simulation, gray markers -- experimental data.
                }
        \label{fig:3D_gixrf_curves}
    \end{figure}

    In this section we consider a periodic 3D nano-columnar structure of Cr,
    manufactured using electron beam lithography~\cite{altissimo2010ebl} on top of a ${\rm SiO}_2$ substrate.
    The structure of the sample is a regular square grid of box-shaped columns
    (see \hyperref[fig:sketch]{Fig.~\ref{fig:sketch}c}) on a substrate,
    with $300~{\rm nm}~\times~300~{\rm nm}$ lateral box dimensions and a $D_x = D_y = 1 {\rm \mu m}$ grid.
    The nominal height of the nanocolumns is $d = 25$~nm.

    GIXRF measurements were carried out at the four crystal monochromator (FCM) beamline~\cite{krumrey2001high}
    in PTB laboratory~\cite{senf1998plane} of the BESSY II storage ring
    and reported by Dialameh et.al.~\cite{dialameh2018development}
    The incident photon energy was $E = 7~{\rm keV}$.
    Numerical simulations are done similarly to those in \autoref{sec:2Dgixrf}.
    The GIXRF experimental data and the best-fit obtained from dynamical diffraction theory simulations are shown in \autoref{fig:3D_gixrf_curves},
    for a selection of azimuthal angles.

    Best-fit model parameters are: lateral period of the structure $Dx = Dy = 1\mu{\rm m}$, matching the same nominal values,
    lateral sizes of the nanocolumns are $300~{\rm nm}~\times~300~{\rm nm}$,
    nanocolumns height $d = 24~{\rm nm}$.
    The best-fit model suggests that there is no surface oxidation $d_t = 0~{\rm nm}$,
    however effective thickness of the side walls is $d_s = 1.3~{\rm nm}$.
    The density of the nanocolumns material is equal to the nominal Cr density $\rho_{\rm Cr} \approx 7.2~{\rm g}/{\rm cm}^3$,
    while substrate density is $\rho_{{\rm SiO}_2} = 2.4~{\rm g}/{\rm cm}^3$.
    Considering the large lateral period $Dx = Dy = 1\mu{\rm m}$ (significantly larger than that of the \grating lamellar grating structure)
    the sidewalls tilt is negligible, therefore $\sigma = 0~{\rm nm}$,
    i.e. the best-fit model for the nano-column structure implies perfectly parallel sidewalls $\langle\chi_h\rangle \equiv \chi_h$.
    % The interpolated experimental GIXRF map and the theoretical GIXRF map are shown in~\autoref{fig:3Dmap}.
    Experimental GIXRF curves \autoref{fig:3D_gixrf_curves} are in good agreement with numerical simulations.

    It is important to note that in the case of grazing incidence geometry,
    the GIXRF curves calculated for the 3D structure could also be approximated with the use of an effective 2D model,
    albeit with reduced density.
    This is because in the grazing incidence geometry the momentum transfer $|k_y| \gg |k_x|$.
    In other words,
    measurements in grazing incidence geometry
    are sensitive to the lower frequencies of the Fourier transform of the structure along the $x$ direction
    and to the higher frequencies along the $y$ direction,
    while the spacing between nodes in reciprocal space along $k_x$ and $k_y$ direction are identical due to the symmetry $D_x = D_y$ of the periodic structure.

    \begin{figure}[t]
        \center\includegraphics[width = \columnwidth]{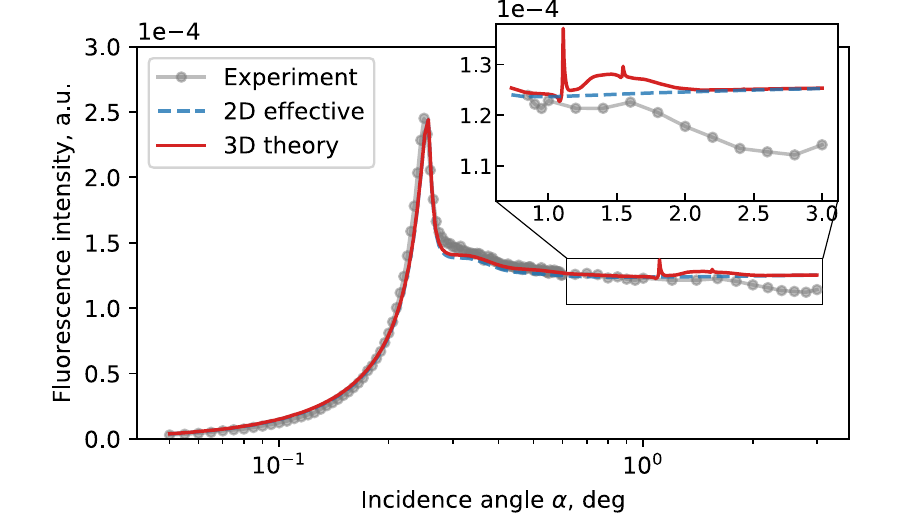}
        \caption{Comparison of effective 2D and genuine 3D simulations of GIXRF Cr-K$\alpha$ curve for 3D Cr nanocolumns structure in conical geometry ($\phi = 0^\circ$).}
        \label{fig:3Dvs2D}
    \end{figure}

    Thus, GIXRF curves of Cr nanocolumns can be effectively represented in a first approach as a lamellar Cr grating with reduced density equal to the averaged density of the actual 3D structure.
    However, a direct comparison between 3D and 2D simulations (figure. \autoref{fig:3Dvs2D}) reveals some differences.
    For higher incident angles above the critical angle of total external reflection, the 2D model (dashed blue line in \autoref{fig:3Dvs2D}) yields a monotonous angular dependence,
    while the experimental GIXRF curve clearly exhibits oscillatory behaviour in that angular range,
    with a maximum at $\alpha \approx 1.5^\circ$.
    In \autoref{fig:3Dvs2D}, curves are shown only for $\phi = 0^\circ$,
    but this oscillation in the range of higher incidence angles $\alpha$ is present in all experimental curves measured at different azimuthal orientations of the sample (see \autoref{fig:3D_gixrf_curves}).
    We attribute this oscillation to interference due to the periodicity of the structure along the $y$ direction,
    which becomes more important at higher incident angles since the value of $|k_y|$ decreases with increasing incidence angle $\alpha$
    and the measurement becomes more sensitive to the lower frequencies of the Fourier transform along the $y$ direction.
    Such interference mode is not taken into account in the 2D simulations.
    Additionally, the 3D simulations show resonant peaks at
    $\alpha \approx 1.15^\circ$ and $\alpha \approx 1.54^\circ$
    which are not resolved in experimental data.
    To observe these peaks,
    measurements with step sizes of $\delta\alpha = 0.01^\circ$ should be resolved,
    which is experimentally feasible, as the resolution limit of modern synchrotron sample stage equipment is on the level of $0.001^\circ$.

\section{Discussion}

    \begin{figure*}[t!]
        \center\includegraphics[width = 1.5\columnwidth]{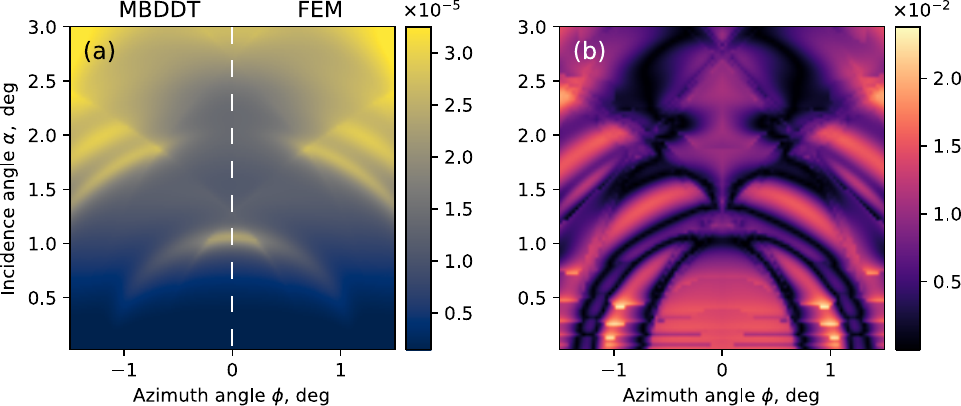}
        \caption{
                    (a) Comparison of GIXRF maps as simulated by MBDDT (left-hand side) and FEM (right-hand side) approaches.
                    (b) Relative discrepancy
                }
        \label{fig:comparison}
    \end{figure*}

    In \autoref{tab:2D_results} we compare the structure parameters of the 2D lamellar \grating grating,
    as reconstructed using the MBDDT simulations described in \autoref{sec:yield},
    with the nominal parameters used in fabrication of the grating.
    The results of of the MBDDT reconstruction are in good agreement with the nominal values.

    To further validate the computational scheme described in~\autoref{sec:yield}, also FEM simulations has been performed.
    The FEM simulations were done using the JCMwave software~\cite{pomplun2007adaptive} for a box model based on the best fit parameters in~\autoref{tab:2D_results}.
    JCMwave is a rigorous Maxwell-solver,
    which enables field simulations in structures of arbitrary shape.
    For the calculation of the finite element solution, the computational domain is meshed into patches where a number of polynomial ansatz functions is defined. The finite element side constraint of 4~nm and a polynomial degree of 4, has been used in the simulations and the GIXRF fluorescence intensities were calculated from electric fields as described in \cite{soltwisch2018element}.
    A direct comparison of the MBDDT and the FEM simulations is shown in~\autoref{fig:comparison}.
    The GIXRF maps are symmetrical with respect to an axis at $\phi = 0^\circ$. In \hyperref[fig:comparison]{Fig.~\ref{fig:comparison}a} the left-hand side map is thus showing the MBDDT result, whereas the right side shows the FEM result. Both simulation results are visually identical. In addition, the relative discrepancy is shown in \hyperref[fig:comparison]{Fig.~\ref{fig:comparison}b}.
    Here, the relative discrepancy is defined as
    $$
    \varepsilon_{ij} = \dfrac{|Y^{\rm (f)}_{ij} - Y^{\rm (m)}_{ij}|}
                             {\max\{Y^{\rm (f)}_{ij}, Y^{\rm (m)}_{ij}\}},
    $$
    where $Y^{\rm(f,m)}_{ij}$ are the GIXRF intensities calculated in each  $(\alpha_i, \phi_j)$ point using the FEM and MBDDT methods respectively.

    The absolute maximum of the relative discrepancy is $2.4\%$ and the discrepancies are generally higher for the low incidence angles. It should be noted, that the precision of the FEM calculation in this angular range may be limited due to the exponential decay of the evanescent waves. In general, the relative discrepancy is on a level of $1\%$ for $80\%$ of the points, proving the validity of the MBDDT approach for such GIXRF simulations.

    In \autoref{tab:3D_results} we compare the MBDDT derived structure parameters of the Cr nanocolumns with their nominal parameters. A good agreement is obtained,
    especially since for the current case only a simple box model is used for the numerical simulations to describe the distribution of fluorescent atoms in the structure.

    \begin{table}[b!]
    \caption{
                Comparison of the 2D structure parameters of the \grating lamellar grating
                as reconstructed by MBDDT with nominal parameters.
             }
    \label{tab:2D_results}
    \center\begin{tabular}{lcc}
    \hline
    \hline
                            & \textbf{Nominal}         &  \textbf{Simulation} \\
    \hline
        Period $D_x$,~nm  & $100$    &   $ 100 $   \\
        Line height $d$,~nm           &  $ 87 $  &  $ 88.7 $   \\
        Line width,~nm    & $ 40 $    &   $ 39 $   \\
        Effective surface thickness $d_t$,~nm      & ---      &  $ 3.3 $   \\
        Effective edge thickness $d_s$,~nm      & ---      &  $ 0 $   \\
        Side walls tilt,~deg  & ---     &  $ 4 $   \\
        Line density $\rho_{{\rm Si}_3{\rm N}_4}$,~$\text{g}/\text{cm}^3$  & $ 3.2 $      &  $ 2.8 $   \\
        Substrate density $\rho_{\rm Si}$,~$\text{g}/\text{cm}^3$  & $ 2.33 $     &  $ 2.22 $   \\
    \hline
    \hline
    \end{tabular}
    \end{table}

    \begin{table}[b!]
    \caption{   Comparison of the 3D structure parameters
                of the Cr nanocolumns
                as reconstructed by MBDDT with nominal parameters.
            }
    \label{tab:3D_results}
    \center\begin{tabular}{lcc}
    \hline
    \hline
                            & \textbf{Nominal}   &  \textbf{Simulation} \\
    \hline
        Period $D_{x,y}$,~$\mu\text{m}$         & $ 1 $  &  $ 1 $   \\
        Column height $d$,~nm           & $ 25 $    & $ 24 $    \\
        Column width,~nm          & $ 300 $  &  $ 300 $   \\
        Effective surface thickness $d_t$,~nm      & ---    & $ 0 $    \\
        Effective edge thickness $d_s$,~nm      & ---    & $ 1.3 $    \\
        Column density  $\rho_{\rm Cr}$,~$\text{g}/\text{cm}^3$  & $ 7.19 $    & $ 7.2 $    \\
        Substrate density  $\rho_{{\rm SiO}_2}$,~$\text{g}/\text{cm}^3$  & $ 2.65 $    & $ 2.4 $    \\
    \hline
    \hline
    \end{tabular}
    \end{table}

    With the MBDDT approach, it is also possible to take into account a structure with tilted sidewalls and surface oxidation.
    The model of the structure would needed to be discretized along the $z$ direction,
    e.g. according to~\cite{pisarenco2016fast}. The sample can be approximated as a stack of homogeneous and/or structured layers,
    where each layer can have arbitrary structure parameters with the exception of the period,
    which must be maintained throughout the whole stack.

    The main benefit in applications of 3D XSW technique to the characterization of nanostructures is its sensitivity to the spatial distribution of the fluorescent atoms within the structure.
    Although the examples considered in this work exhibited homogeneous lateral distribution of N and Cr within the grating line and nanocolumn,
    we can still demonstrate this sensitivity by performing simple calculations.

    \begin{figure}[b!]
        \center\includegraphics[width = 1\linewidth]{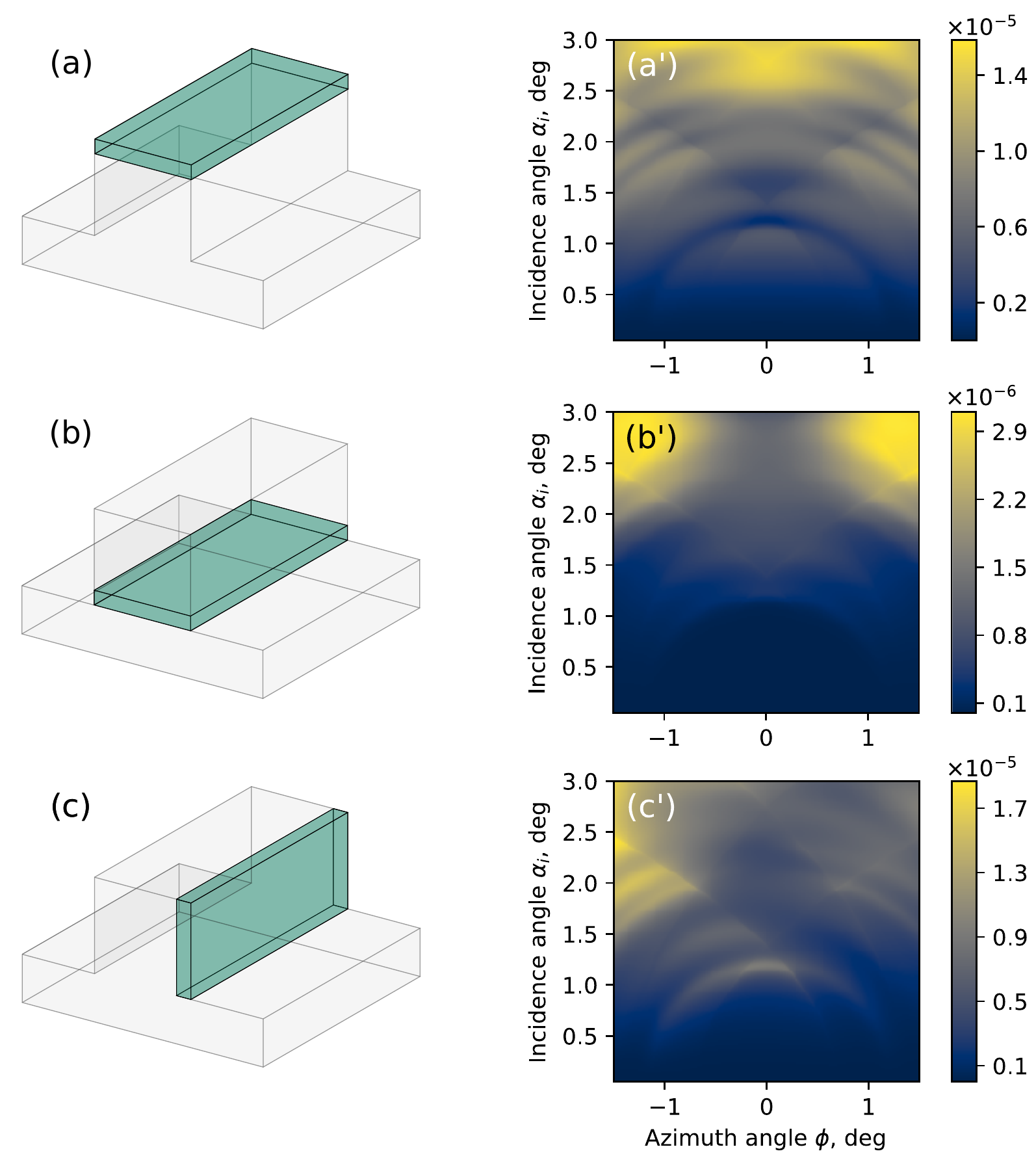}
        \caption{
                    Simulation of GIXRF maps for inhomogeneous distribution of fluorescent atoms within the lamellar grating structure.
                    From (a) to (c): sketch of the structure,
                    green box depicts the localization of the dopant atoms.
                    From $({\rm a}')$ to $({\rm c}')$: corresponding GIXRF maps.
                }
        \label{fig:inhomogenious_distribution_hort}
    \end{figure}

    We use the same model for the \grating lamellar grating as already shown earlier,
    but now we assume to have a  dopant atoms to be localized in a confined volume within the structure as shown in \hyperref[fig:inhomogenious_distribution_hort]{Figs.~\ref{fig:inhomogenious_distribution_hort}(a--c)} instead of being homogeneously distributed,
    as assumed in \autoref{sec:2Dgixrf}.
    The specific localization of the dopant atoms is depicted as green boxes and the resulting simulated GIXRF maps for the calculated fluorescence signal of the dopant atoms are shown.
    It can be observed that the corresponding GIXRF maps are highly sensitive to this variation. For an asymmetric distribution of fluorescent atoms
    \hyperref[fig:inhomogenious_distribution_hort]{Fig.~\ref{fig:inhomogenious_distribution_hort}c},
    also an asymmetry is observed in the GIXRF maps. One may exploit such asymmetry,
    e.g. to distinguish chemical compositions of the left and right sidewall of the grating line.
    This may be useful in, e.g., the characterization of gratings fabricated with multi-patterning~\cite{weber2012high} techniques.

\section{Conclusions}

     A new computational scheme based on the dynamical diffraction theory has been developed and applied for the analysis of GIXRF experiments on 2D and 3D periodic nanostructures.
     It is capable of simulating GIXRF data from structures with specific element distributions both in-plane as well as in-depth.
     The computational scheme has been validated with a Maxwell solver based on the finite element method and benchmarked on GIXRF experimental data obtained from \grating 2D lamellar gratings and Cr 3D nanocolumns.
     The reconstructed geometrical parameters of the lamellar grating derived from the elemental distribution are in good agreement with nominal values,
     as well as with parameters obtained from a previous study performed using a finite element method.
     Furthermore, the parameters of the elemental distribution in the Cr 3D nanocolumns were reconstructed for the first time. A reconstruction of the geometrical parameters of this structure by means of FEM is practically impossible due to the required higher excitation photon energy, the larger period of the structures (and thus larger computational cell) and the 3D dimensionality of the sample. The obtained results of this reconstruction are in good agreement with the nominal.
     Finally, we conclude that the MBDDT computational scheme can be used in conjunction with the GIXRF experimental technique as a powerful tool in element selective nanometrology for 2D and 3D periodic structures.

\begin{acknowledgments}
    This work is part of the research programme of the Industrial Focus Group XUV Optics, being part of the MESA+ Institute for Nanotechnology and the University of Twente (www.utwente.nl/xuv).
    It is supported by ASML, Carl Zeiss SMT AG and Malvern Panalytical, as well as the Province of Overijssel and the Netherlands Organization for Scientific Research (NWO).
    This project has received funding from the Electronic Component Systems for European Leadership Joint Undertaking under grant agreement No 826589 — MADEin4.
    This Joint Undertaking receives support from the European Union's Horizon 2020 research and innovation programme and Netherlands, France, Belgium, Germany, Czech Republic, Austria, Hungary, Israel.

\end{acknowledgments}

\bibliographystyle{apsrev4-1}
\bibliography{bibliography}

\end{document}